\documentclass{ws-procs9x6}

\def\cal{\mathcal}  %WHY won't it use calligraphics?

\def\lra{\leftrightarrow}
\def\x{{\bm x}}
\def\p{{\bm p}}
\def\k{{\bm k}}
\def\q{{\bm q}}

\def\h{{\bm h}}

\def\F{{\bm F}}

\def\Re{{\rm Re}}

\def\twotwo{{2\lra2}}

\def\half{{\textstyle\frac{1}{2}}}

\def\ca{C_{\rm A}}
\def\cf{C_{\rm F}}

\def\df{d_{\rm F}}

\def\mD{m_{\rm D}}
\def\mg{m_{{\rm eff,g}}}

\def\mq{m_{{\rm eff,q}}}

\def\n{{\hat{\bm n}}}

\def\degen{\nu}

\def\splitsym{\gamma}

\def\slashchar#1{\setbox0=\hbox{$#1$}           % set a box for #1 
   \dimen0=\wd0                                 % and get its size
   \setbox1=\hbox{/} \dimen1=\wd1               % get size of /
   \ifdim\dimen0>\dimen1                        % #1 is bigger
      \rlap{\hbox to \dimen0{\hfil/\hfil}}      % so center / in box
      #1                                        % and print #1
   \else                                        % / is bigger
      \rlap{\hbox to \dimen1{\hfil$#1$\hfil}}   % so center #1
      /                                         % and print /
   \fi}                                         %

\begin{document}

\title{Transport Coefficients at Leading Order:  Kinetic Theory versus
Diagrams}

\author{G.~D.~Moore}

\address{Department of Physics, \\ McGill University, \\
	3600 rue University, \\ Montreal, QC H3A 2T8, Canada \\
	E-mail:  guymoore@physics.mcgill.ca}

\maketitle

\abstracts{I review what is required to compute transport coefficients
in ultra-relativistic, weakly coupled gauge theories, at leading order
in $g$, using kinetic theory.  Then I discuss how the calculation would
look in alternative approaches:  the 2PI method, and direct diagrammatic
analysis.  I argue that the 2PI method may be a good way to derive the
kinetic theory, but is not very useful directly (in a gauge theory).  
The diagrammatic approach is almost hopeless.}

\section{Introduction}

A ``big question'' in particle physics at extreme energy
densities (Strong and Electroweak Matter, the topic of this conference),
is how nonequilibrium systems behave.  Here I will only talk about
weak coupling, plus some conditions on the departure from equilibrium
and the measurables of interest, which allow application of 
techniques grounded in perturbative field theory.
(Some very interesting questions are excluded by this condition; but the
problems included are interesting and challenging.)  In particular, I am
going to talk about transport 
coefficients, even though they represent the ``closest to equilibrium''
nonequilibrium systems we could consider.

Transport coefficients tell about equilibration in systems which are
homogeneous on fairly large scales, and are locally fairly close to
equilibrium.  The traditional approach to dealing with such a system at
weak coupling is kinetic theory, that is, Boltzmann equations.  We have
recently presented\cite{AMY1} a Boltzmann equation which is
adequate for treating transport coefficients in a nonabelian gauge
theory at leading order in the coupling $g$ (see also Larry Yaffe's talk
in this proceeding).  Its applicability extends
well beyond the near-equilibrium situation needed for transport, though
the homogeneity requirements for the equation to apply are stronger than
you might naively guess.

But there has been much interest in addressing transport using other
tools, particularly the 2PI formalism and direct diagrammatic analysis.
I will argue that, if the goal is to perform a calculation in a gauge
theory, which is complete at leading order in the gauge coupling $g$,
then both methods become very complicated.  The 2PI formalism may be a
very efficient way of deriving the correct Boltzmann equation, but
direct application without gradient expansion--for instance,
numerically, as has recently been done in scalar field models%
\cite{Berges1,Berges2,Berges3}--will encounter substantial
difficulties.  The diagrammatic approach at leading order appears to be
so complicated that I see little reason to pursue this method further.

\section{Boltzmann approach to transport coefficients}
\label{sec:transport}

When a plasma is sufficiently homogeneous, the large scale flow is
accurately described by ideal hydrodynamics.  Under ideal hydrodynamic
behavior, entropy is conserved; particle number is often approximately
conserved.  In the heavy ion context, this means that almost all 
the available energy of the system goes into bulk flow.  

Transport coefficients give the first 
corrections to this behavior, when the system is large but not
infinitely large.  They cause entropy generation,
and tell how close to local thermal equilibrium the system remains.  In the
context of heavy ion collisions, knowing the transport coefficients
could tell us when the hydrodynamic approximation has broken
down, which would help hydro simulators to know when to impose chemical
and kinetic freeze-out.  In early universe physics,
diffusion coefficients are useful in baryogenesis calculations, and
viscosity and diffusion coefficients can indicate how far around
electroweak bubble walls the departure from equilibrium extends.  For
the ``pure'' theory of thermal field theory, transport coefficients are
a useful object of study because their definition is very theoretically
clean.  If we want to develop the tools to study nonequilibrium problems
in a controlled way, those tools should be able to tell us the values of
transport coefficients.  So being able to compute them is the most basic
step towards more general nonequilibrium calculations.

In the Boltzmann approach, one shows (or assumes, or hopes) that the
plasma is well described as a bath of long lived
quasiparticles, which undergo occasional and relatively localized
scatterings.  That is, one requires that the interesting observables are
dominated by the two point function, that the spectral weight for the
two point function is tightly peaked, that the evolution of the two
point function can be accurately written in terms of scatterings with
the other quasiparticles, and that these scattering processes are local
on the scale of the spatial variation of the system.

Given these conditions, one can write down a Boltzmann equation,
\begin{equation}
\label{eq:Boltzmann}
\partial_t f(\p,\x,t) + \frac{\p}{p^0} \cdot \partial_{\x} f(\p,\x,t)
	+ {\bm F} \!\cdot \partial_\p f(\p,\x,t)
	= - {\cal C}[f] \, ,
\end{equation}
with ${\cal C}$ the collision operator.  Applying this to a situation
where $f$ is spatially inhomogeneous, or where the force term represents
an electrical field, allows calculation of transport coefficients.
In particular, when $f$ exhibits shear flow, 
then computing the traceless part of
the stress tensor gives the shear viscosity, 
\begin{equation}
\label{eq:def_eta}
T_{ij}[f] - \frac{1}{3} \delta_{ij} T_{kk}[f] 
	= \eta \left( \partial_i v_j + \partial_j v_i - \frac{2}{3}
	\delta_{ij} \partial_k v_k \right) \, .
\end{equation}
Similarly, the diffusion constant is determined by the size of a
conserved current, when the number density varies in space.  In each case,
one must solve for $f$ (at linearized order in the departure from local
equilibrium) in the Boltzmann equation.  This is hard because it is an
integral equation; the operator ${\cal C}[f]$ contains $f$ at momenta
other than $\p$.  It is especially hard because, at leading order,
${\cal C}$ turns out to be quite nasty!

The collision operator ${\cal C}$ should include ``all collision
processes which are important at leading order.''  At leading {\em log}
order, there are only a few processes; $t$ channel boson exchange,
Compton scattering, and an annihilation process related to Compton by
crossing.  Further, simplifying approximations can be made to each
collision term, which in fact makes the collision operator a
differential operator, not an integral one\cite{AMY1}.  But at leading
order, in fact at next-to-leading-logarithm, one needs not only all 
$2 \leftrightarrow 2$ processes, but certain inelastic LPM suppressed
splitting processes\cite{AMY5}.  For the full details see%
\cite{AMY5,AMY2,AMY4}.  To show how bad the situation is,
I will write ${\cal C}$ in some detail,
though not enough to allow you to evaluate it:  The 
$2 \leftrightarrow 2$ piece is
\begin {eqnarray}
   {\cal C}^\twotwo_a[f]
  & = & \frac 1{4|\p|\degen_a} \sum_{bcd} \int_{\k\p'\k'} \!\!
       \left|{\cal M}^{ab}_{cd}(\p,\k;\p',\k')\strut\right|^2 \!
       (2\pi)^4 \delta^{(4)}(P{+}K{-}P'{-}K')
\nonumber\\ && \times
       \Bigl\{
          f_a(\p) \, f_b(\k) \, [1{\pm} f_c(\p')] \, [1{\pm} f_d(\k')]
          - [ (p,k) \leftrightarrow (p',k') ]
       \Bigl\} \,.
\label{eq:Ctwotwo}
\end {eqnarray}
The $2 \leftrightarrow 2$ matrix elements should be summed over incoming
and outgoing spins and colors; literature values, summed only over
outgoing states, are available\cite{Combridge}.  But in addition,
HTL corrections are needed on some internal lines, which means that
the form of certain matrix elements is more complicated than in
vacuum\cite{AMY5}.  But the real problem is the inelastic LPM suppressed
splitting processes: 
\begin {eqnarray}
   {\cal C}^{``1\lra2"}_a[f]
   & = &
       \frac{(2\pi)^3}{2|\p|^2 \degen_a}
       \sum_{b,c}
       \int_0^\infty dp' \> dk' \;
       \delta(|\p|-p'-k') \;
       \splitsym^a_{bc}(\p;p'\, \hat \p,k'\, \hat \p)
\nonumber\\ && \hspace{4em} \times
       \Bigl\{
          f_a(\p) [1\pm f_b(p' \, \hat \p)] [1\pm f_c(k' \, \hat\p)]
          - [ f \leftrightarrow (1{\pm}f)]
       \Bigr\}
\nonumber\\ &&
    +
       \frac{(2\pi)^3}{|\p|^2 \degen_a}
       \sum_{b,c}
       \int_0^\infty dk \> dp' \;
       \delta(|\p|+k-p') \;
       \splitsym_{ab}^{c}(p'\, \hat\p;\p,k \,\hat\p)
\nonumber\\ && \hspace{4em} \times
       \Bigl\{
          f_a(\p) f_b(k \, \hat\p) [1\pm f_c(p' \, \hat\p)]
          - [ f \leftrightarrow (1{\pm}f)]
       \Bigr\} \, .
\label {eq:C12form}
\end {eqnarray}
The awfulness is hiding in $\gamma^a_{bc}(\ldots)$; evaluating it requires
solving an integral equation.  For instance, for $q \leftrightarrow qg$
processes, it is
\begin{eqnarray}
\gamma^q_{qg}(p \n; p' \n, k \n)
   &=&
	\frac{p'{}^2 + p^2}{p'{}^2 \, p^2 \, k^3} \>
	{\cal F}_{\rm q}^\n(p,p',k) \,,
	\nonumber \\
{\cal F}_{\rm q}^\n(p,p',k) & \equiv & \frac{\df \cf \alpha_{\rm s}}
	{2(2\pi)^3} \int \frac{d^2 \h}{(2\pi)^2} \;
	2 \h \cdot \Re {\bf F}_s^\n(\h;p,p',k) \, ,
	\nonumber \\
	2\h & = &
    i \, \delta E \, \F_s^\n(\h)
        + g^2 T \int \frac{d^2 \q_\perp}{(2\pi)^2}
	\left( \frac{1}{\q_\perp^2} - \frac{1}{\q_\perp^2 + \mD^2}
	\right)
\nonumber\\ && \hspace{0.5in} \times
	\biggl\{
	     (C_s - \half \ca) \left[{\F}_s^\n(\h)
                       - {\F}_s^\n(\h{-}k\,\q_\perp) \right]
\nonumber\\ && \hspace{1.05in}
	     + \half \ca \left[{\F}_s^\n(\h)
                       - {\F}_s^\n(\h{+}p\,\q_\perp) \right]
\nonumber\\ && \hspace{1.05in}
	     + \half \ca \left[{\F}_s^\n(\h)
                       - {\F}_s^\n(\h{-}p'\q_\perp) \right]
	\biggr\} %\biggr]
	\nonumber \\	
    \delta E
    & = &
	\frac{\mg^2}{2k} + \frac{\mq^2}{2p'} - \frac{\mq^2}{2p} 
	+ \frac{\h^2}{2p \, k\, p'} \, .
\label{eq:horror}
\end{eqnarray}
That is, there is an integral equation just to determine the coefficient
in one of the collision terms, which itself appears in an integral
equation for the expression that we want.

\section{Alternate approaches}

For some reason, people don't like the Boltzmann approach.  The obvious
reason is that it involves solution of an integral equation (in fact, at
leading order, an integral equation within an integral equation).
This objection is probably not well grounded,
because other approaches presumably reduce to the 
same integral equation, in one guise or another.
But there are some deeper, more philosophical reasons people don't like
the Boltzmann approach.  

One problem is that people don't trust it.  Where is the careful
derivation?  How do you know you have not missed some necessary term in
the collision integral?  What about quantum and high gradient effects?
What do you do when the particle widths are large?  Another problem
people have is that the Boltzmann approach is old fashioned.  Boltzmann
was playing with such equations a hundred years ago.  Surely there is a
more powerful modern approach.

\subsection{The 2PI approach}

One approach much advocated in the recent literature%
\cite{CalzettaHu,CalzettaHu2,Berges1,Berges2,Berges3} (see also
contributions to this proceedings from Berges and from Mottola) is the
2PI approach.  The idea, in summary, is to
solve for the evolution of the 2-point function by variationally
minimizing a functional $\Omega(\Delta)$ of the 2-point function
$\Delta$.  This functional looks 
like a logarithm of the 2-point function, plus a term involving the
propagator and self-energy, plus $\Phi$, the sum of 2
particle irreducible (2PI) bubble diagrams; inside the functional the 2-point
function is treated as arbitrary.  If the complete set of 2PI diagrams
is included, the method is exact; but generally, to get anywhere
it is necessary to truncate in some way the sum over 2PI diagrams.
There are two approaches to using the 2PI formalism 
which particularly recommend themselves:
\begin{enumerate}
\item 
Make additional approximations, in particular, make a gradient expansion
and assume slow time development.  Similar approximations to those
Calzetta and Hu used in scalar theory\cite{CalzettaHu} may be the most
efficient way to derive the Boltzmann equation described in the last
section.  
\item
Take no prisoners.  Directly solve the evolution of the 2-point
function, probably numerically on a lattice.  This has been done for
scalar field theory in 1+1 dimensions\cite{Berges1,Berges2}, and more
recently in 3+1 dimensions within a large $N$ expansion\cite{Berges3}.
\end{enumerate}

\begin{figure}[h]
\centerline{\epsfxsize=4in\epsfbox{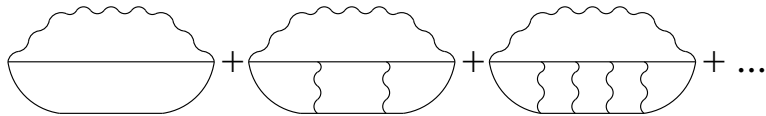}}
\caption{Some QED diagrams needed at leading order. \label{fig:qed}}
\end{figure}

The 2PI approach is much trickier in a gauge theory
than in a scalar theory, where most applications have been made.  In a
scalar theory it is generally assumed that the importance of a diagram,
in the $\Phi$ functional, is determined by its loop order.  But this is
not completely obvious.  And in a gauge theory, it is not true.  The
problem is that there are special kinematic ranges in a gauge
theory, namely soft momenta and collinear momenta, where enhancements
occur which can obviate naive loop counting.  For instance, in QED, I
claim that all the diagrams shown in Fig.\ref{fig:qed} 

\noindent
are important at leading order.
The proof that these are leading order is to open the photon line, to get
a photon self-energy:
\begin{figure}[h]
\centerline{\epsfxsize=4.5in\epsfbox{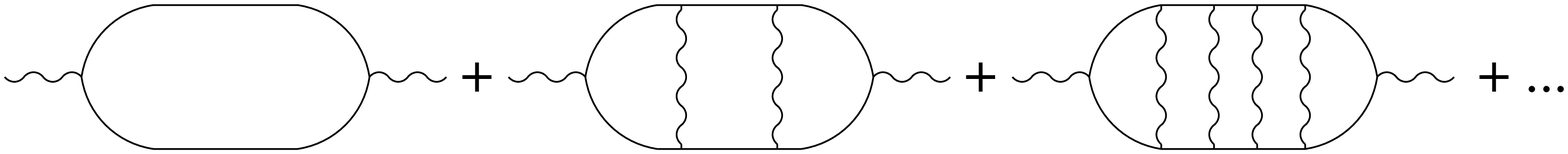}}
\caption{Same diagrams, with photon line opened. \label{fig:qed2}}
\end{figure}

\noindent
and then to read our paper\cite{AMY2}, which does the detailed power
counting.  

\begin{figure}[h]
\centerline{\epsfxsize=4.5in\epsfbox{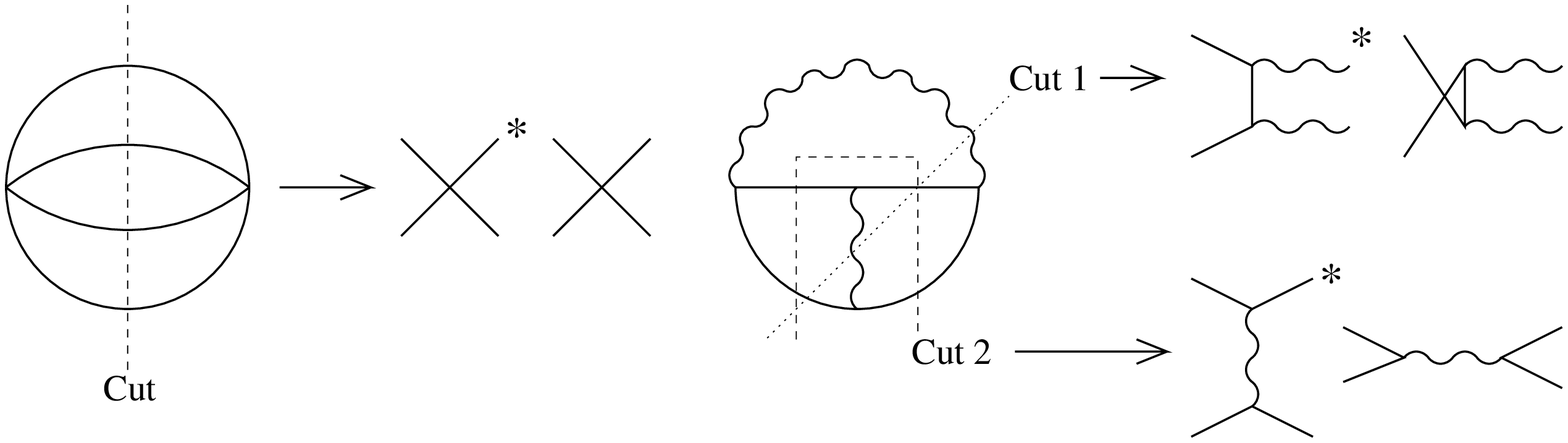}}
\caption{Left:  scalar 2PI diagram which gives $2 \leftrightarrow 2$
scattering.  Right:  QED diagram which gives an interference term
between scattering amplitudes, in more than one way. \label{fig:cuts}}
\end{figure}

Cutting a 2PI diagram gives a matrix element and its conjugate.  Already
knowing what scattering matrix elements we need for the Boltzmann
equation, we can work backwards to figure out what 2PI diagrams must be
required.  For instance,
a 3-loop diagram leads to interference effects which are important at
leading order in QED, as shown in Fig.~\ref{fig:cuts}.

\begin{figure}[h]
\centerline{\epsfxsize=3.5in\epsfbox{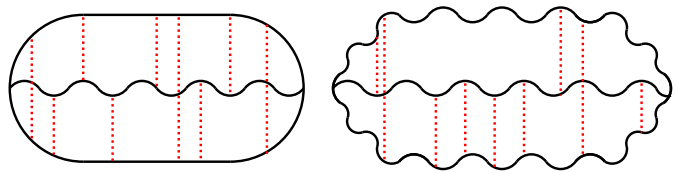}}
\caption{Examples of required diagrams. \label{fig:horror2}}
\end{figure}

The trouble is that the inelastic, LPM suppressed splitting processes
arise from the sum over an infinite set of diagrams; for instance,
the interference between two of the amplitudes for bremsstrahlung with 3
\begin{figure}[h]
\centerline{\epsfxsize=3.5in\epsfbox{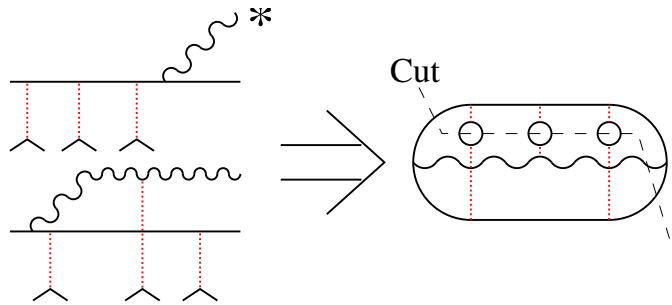}}
\caption{Two processes whose interference is important in determining
the gluon bremsstrahlung rate, and the 2PI diagram they imply.  Red
lines are soft ($gT$) gauge propagators in the Landau cut.
\label{fig:horror1}}
\end{figure}
scatterings is shown in Fig.~\ref{fig:horror1}.  In this and
following figures, dotted lines are soft spacelike gauge boson propagators
(in the Landau cut), while solid lines are always hard and on-shell; I
use wavy rather than curly 
lines for gluons to make the pictures less cluttered.
I have shown the loops responsible for the spectral weight of the soft
lines for clarity; all lines in the 2PI formalism are always self-energy 
resummed.  This is one of a family of diagrams which must be included.
Any number of soft lines are permitted, provided that they do not
cross; so for instance, in the calculation of shear viscosity it is
necessary to include 2PI diagrams such as those shown in
Fig.~\ref{fig:horror2}, {\em just to get leading order results in }
$g_{\rm s}$.

If one is willing to make extra approximations, the presence of this
infinite class of diagrams need not be fatal.  There is a parametric
difference between the soft exchange momenta and the hard momenta of the
main lines in the diagrams.  The soft exchanges can be resummed.  This
leads to the integral equation for the splitting process 
already presented in Eq.~(\ref{eq:horror}); together with a
gradient expansion, this should be the most efficient way to derive the
Boltzmann equation presented in the last section.  However, for the 
take-no-prisoners approach, an infinite set of diagrams is a
show-stopper.  Either some approximation must be made to resum them (and
this probably demands some gradient expansion treatment or separation of
time scales), the set of diagrams must be truncated, or the numerical
approach becomes impossible.  Truncating the set of diagrams raises new
problems, because it is only the whole set, not any truncation, which is
gauge invariant, and even then only when the distinction between $gT$
and $T$ momenta is made parametric.  
(A truncation also obviously means abandoning a leading order
treatment--in fact, even a next to leading log treatment.)

In conclusion, the 2PI formalism looks like a good way to derive the
Boltzmann equation.  It may provide a good framework for a consistent
power counting effort, to prove that the claimed set of diagrams is
complete.  But as a direct computational tool, it appears ill suited in
a gauge theory.

\subsection{Diagrammatic approach}

\begin{figure}[h]
\centerline{\epsfxsize=4in\epsfbox{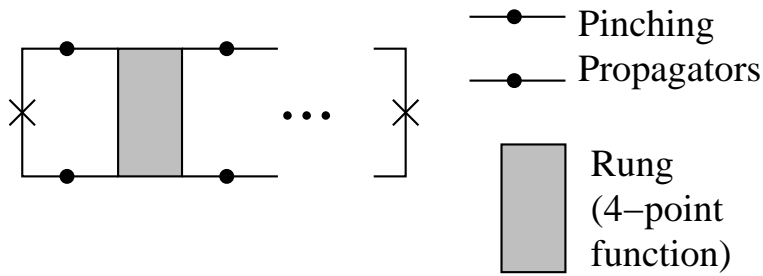}}
\caption{General structure of ladder diagrams.  Diagrams with any number
of rungs (0,1,2,3,$\ldots$) are needed.  The crosses are operator 
($T_{\mu \nu}$ or current) insertions.  \label{fig:ladder}}
\end{figure}

The diagrammatic approach to evaluating transport coefficients, in
relativistic field theory, was pioneered by Jeon\cite{Jeon}.  Recently
his (inefficient) treatment has been streamlined\cite{Heinz,Basagoiti}
and applied, at leading logarithmic order, to gauge theory%
\cite{Basagoiti}.  Jeon showed that the diagrams which contribute at
leading order are ladder diagrams, with a general form shown in 
Fig.~\ref{fig:ladder}. 

The self-energy needed on the pinching, nearly on-shell propagators is
determined by differentiating the $\Phi$ functional with respect to the
propagator once (opening one
line), while the rung is obtained by differentiating twice (opening two
lines).  In scalar field theory, this leads to a relatively tidy set of
diagrams.  But in a gauge theory, the very complicated 2PI diagrams
discussed above lead to a set of diagrams, required at leading order,
which is rather intimidating.  For instance, 
\begin{figure}[h]
\centerline{\epsfxsize=3in\epsfbox{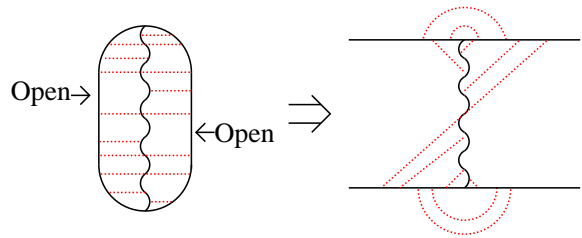}}
\caption{A bubble, which when opened gives a horrible ladder rung.
\label{fig:horror3}}
\end{figure}
in Fig.~\ref{fig:horror3} I show the result of opening one of the
required 2PI bubble graphs on two of its lines, to get a required ladder
rung.  

\begin{figure}[h]
\centerline{\epsfxsize=4in\epsfbox{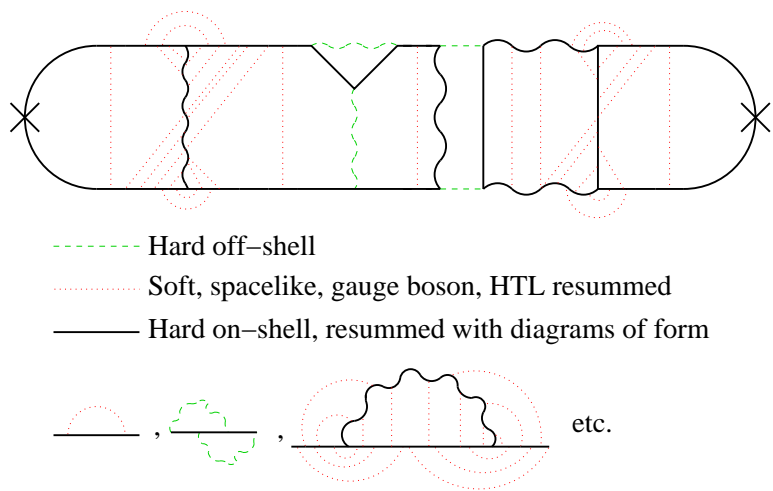}}
\caption{A typical diagram needed in the calculation of leading order
shear viscosity in QCD. \label{fig:shear_diagram}}
\end{figure}

An illustrative example of one of the graphs in the general class
which must be resummed, to determine the shear viscosity of QCD with
fermions, is shown in Fig.~\ref{fig:shear_diagram}.
This diagram will hopefully give the impression that computing transport
at leading order in $g_{\rm s}$, in ultra-relativistic gauge theories,
by a direct diagrammatic attack, is pretty hopeless.

\section{Conclusions}

Don't get me wrong.  I actually quite like the 2PI approach to
dynamics.  But it is not well suited to gauge theories,
except as a method to do a convincing derivation of
the Boltzmann equations, which should then be used to solve problems
like transport coefficients.  Direct numerical application on a lattice,
such as has been done in scalar theories\cite{Berges1,Berges2,Berges3},
is probably hopeless (if complete leading order in $g_{\rm s}$ results
are desired--a Boltzmann-Vlasov approach, valid for 
transport at leading log order, may be feasible and might have some
utility). 

As for the diagrammatic approach:  I defy its supporters to give a
convincing power counting argument, without reference either to 2PI or
to kinetic theory, that the set of diagrams which I describe above, is
both necessary and sufficient.  Their resummation is expected to be
technically challenging, and in the end it will only
boil down to a derivation of the Boltzmann equations.  I don't see
the benefit of doing this exercise.

\end{document}